\documentclass[conference]{IEEEtran}
\IEEEoverridecommandlockouts
\usepackage{cite}
\usepackage{amsmath,amssymb,amsfonts}
\usepackage{algorithmic}
\usepackage{graphicx}
\usepackage{textcomp}
\usepackage{xcolor}
\def\BibTeX{{\rm B\kern-.05em{\sc i\kern-.025em b}\kern-.08em
    T\kern-.1667em\lower.7ex\hbox{E}\kern-.125emX}}
\begin{document}

\title{Electrochemical Communication in Bacterial Biofilms: A Study on Potassium Stimulation and Signal Transmission
}

\author{\IEEEauthorblockN{Nithin V. Sabu\IEEEauthorrefmark{1}, and
Bige Deniz Unluturk\IEEEauthorrefmark{2}}
\IEEEauthorblockA{\IEEEauthorrefmark{1}Institute of Quantitative Health Science and Engineering,
Michigan State University, East Lansing, MI 48824 USA}
\IEEEauthorblockA{\IEEEauthorrefmark{2} Electrical and Computer Engineering, and Biomedical Engineering,
Michigan State University, East Lansing, MI 48824 USA}
\IEEEauthorblockA{vettikk1@msu.edu\IEEEauthorrefmark{1}, unluturk@msu.edu\IEEEauthorrefmark{2}}
}

%

\newcommand{\Gex}{G_\mathrm{e}}
\newcommand{\Kex}{K_\mathrm{e}}
\newcommand{\Gin}{G_\mathrm{i}}
\newcommand{\Ginput}{G_\mathrm{input}}
\newcommand{\Kinput}{K_\mathrm{input}}
\newcommand{\Gintr}{G_\mathrm{intr}}
\newcommand{\Kin}{K_\mathrm{i}}
\newcommand{\Kintr}{K_\mathrm{intr}}
\newcommand{\Kac}{K_\mathrm{ac}}
\newcommand{\Vthresh}{V_\mathrm{t}}
\newcommand{\Gmax}{G_\mathrm{m}}
\newcommand{\Kmax}{K_\mathrm{m}}
\newcommand{\Gsup}{G_\mathrm{sup}}
\newcommand{\inter}{\mathrm{intr}}
\newcommand{\Mgrow}{M_\mathrm{g}}
\newcommand{\ie}{i.e.,}
\newcommand{\mm}{\text{mm}}
\newcommand{\ms}{\text{ms}}
\newcommand{\mM}{\text{mM}}
\newcommand{\mV}{\text{mV}}
\newcommand{\hr}{\text{hr}}
\newcommand{\tp}{T_\mathrm{p}}
\maketitle

\begin{abstract}
Electrochemical communication is a mechanism that enables intercellular interaction among bacteria within communities. Bacteria achieves synchronization and coordinates collective actions at the population level through the utilization of electrochemical signals. In this work, we investigate the response of bacterial biofilms to artificial potassium concentration stimulation. We introduce signal inputs at a specific location within the biofilm and observe their transmission to other regions, facilitated by intermediary cells that amplify and relay the signal. We analyze the output signals when biofilm regions are subjected to different input signal types and explore their impact on biofilm growth. Furthermore, we investigate how the temporal gap between input pulses influences output signal characteristics, demonstrating that an appropriate gap yields distinct and well-defined output signals. Our research sheds light on the potential of bacterial biofilms as communication nodes in electrochemical communication networks.
\end{abstract}


\section{Introduction}
Bacterial biofilms, which are collections of microorganisms that adhere to surfaces, are well known for their capacity for cooperation and coordination. Quorum sensing, in which bacteria produce and detect signaling molecules to adjust their gene expression in response to changes in population density, is a common method for achieving this collective behavior. While quorum sensing as a communication mechanism has been the subject of much research, electrochemical communication is now being discovered to have a different level of complexity. Bacteria within biofilms use sophisticated electrochemical techniques to transmit information, ensuring their survival and adaptability.

The process of electrochemical communication facilitates intercellular communication among bacteria within communities by employing electrochemical signals to synchronize collective action at the population level. The phenomenon of electrochemical communication in bacteria can be classified into three distinct categories\cite{lee2017}: 1) Diffusive electron transfer facilitated by redox-active molecules, 2) long-range electrochemical communication achieved through the propagation of potassium waves, and 3) short-range direct contact via membrane-associated cytochromes or nanowires. While there has been research \cite{liu2017a,kang2018} on the molecular communication aspect of electron transfer via redox-active molecules, the exploration of communication utilizing nanowires and electrochemical signals remains unexplored. 
This study centers on the investigation of electrochemical  communication in bacterial organisms.

The cell membrane exhibits selective permeability to ions, molecules, and other substances. Pore-forming membrane proteins called ion channels allow the propagation of ions according to the electrochemical gradient. The exact role of ion channels in bacteria had been unresolved until recently. The work \cite{liu2015} has shown that the variation in the nutrient availability causes \textit{Bacillus subtilis} biofilms to undergo metabolic oscillations. 
The study conducted in \cite{prindle2015} revealed that the presence of ion channels in bacteria facilitates the transmission of long-distance electrical signals from the interior to exterior of the biofilms. These electrical signals are propagated through potassium waves, and they are responsible for the observed metabolic oscillations. The authors of \cite{martinez-corral2019} also observed that a potassium signal input at the periphery of the biofilm causes a potassium wave to travel inwards.

Several studies have mathematically characterized the metabolic oscillations, and electrochemical  communication in bacteria \cite{martinez-corral2018,martinez-corral2019,ford2021}. The study conducted in \cite{martinez-corral2018} employed both analytical and numerical methods to investigate metabolic oscillations. Their research utilized a minimal delay-differential equation model and revealed that these oscillations emerge as a result of a subcritical Hopf bifurcation. This bifurcation gives rise to a bistable regime characterized by the coexistence of an oscillating state and a non-oscillating state. The authors of \cite{martinez-corral2019} introduced a spatially extended model in their study, which integrates glutamate metabolism and potassium wave propagation. Their findings demonstrate that this model successfully replicates the experimental data of biofilm oscillations. \cite{ford2021} presented several modifications to the model. These modifications encompassed the following aspects: the response of cells to fluctuations in potassium levels rather than absolute potassium levels, the adoption of a Neumann flux boundary condition at the biofilm interface as opposed to an artificial flux approximation, the incorporation of potassium uptake by cells through leak gates, and the formulation of a two-dimensional model capable of capturing spatiotemporal oscillatory patterns and the interplay between distinct biofilms within a flow-cell.

Our study investigates the electrochemical  communication inside bacterial biofilms from the perspective of molecular communication. We also evaluates the feasibility of utilizing bacterial biofilms as molecular communication nodes based on electrochemical  communication, a topic that has not yet been addressed in existing research. In contrast to the studies conducted in \cite{prindle2015,martinez-corral2018, martinez-corral2019,ford2021}, which mainly investigated naturally occuring  metabolic oscillations and electrochemical communication in bacterial biofilms, our research has a different focus. We aim to examine the electrochemical communication in biofilms when they are artificially stimulated at the interior of the biofilm. Such internal stimulations can amplify and relay the signal outwards to the biofilm's periphery. This can initiate a cascading amplify-and-relay mechanism in adjacent biofilms, facilitating longer-distance signal transmission.  The input signal for this stimulation is an impulse or pulse of potassium concentration. According to our findings, the introduction of a signal at a specific site within the bacterial biofilm results in the transmission of the signal to other regions of the biofilm. This transmission is facilitated by intermediary cells that both amplify and relay the signal.

 The study conducted in \cite{liu2017} revealed that the metabolic oscillations of two distinct biofilm communities of \textit{Bacillus subtilis} can be synchronized through the process of electrochemical  communication. Biofilms undergo a transition from synchronous to alternating oscillations in response to variations in nutrient availability. The potential for establishing a long-distance relay network by leveraging the cooperative nature of numerous biofilms offers a promising avenue for the efficient and dependable conveyance of information between far endpoints. Hence, it is essential for researchers to give attention to the investigation of electrochemical communication in bacteria. The contributions of this work are given below.

\begin{itemize}
\item Modification of existing mathematical models to incorporate glutamate nutrient supply and input signals, encompassing both impulse and pulses of potassium signal.
\item Verification of continuous glutamate nutrient supply as a means to mitigate metabolic stress and suppress natural bacterial metabolic oscillations.
\item Analysis of the channel response when the interior biofilm regions are subjected to impulse and pulse emissions of potassium input signals. Additionally, exploration of the impact of input potassium signals on biofilm growth.
\item Investigation of how the temporal gap between input pulses influences output pulse characteristics. An inadequate gap may attenuate the output response, while a suitable gap yields distinct and well-defined output signals.
\end{itemize}
 
\section{Electrochemical Communication in Bacteria}\label{sec:ECinBac}
\begin{figure}
    \centering
    \includegraphics[width=1\linewidth]{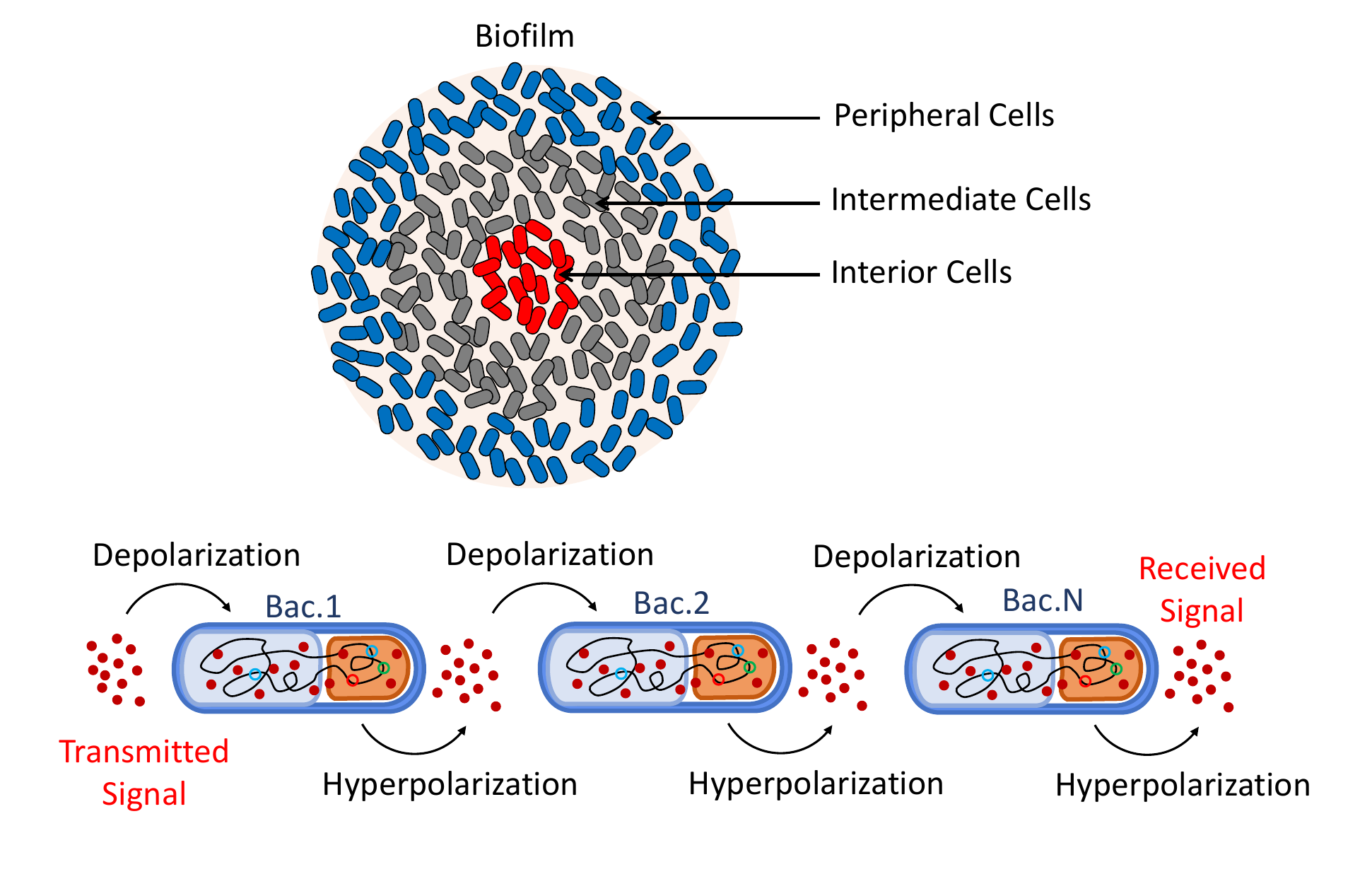}
    \caption{Signal propagation in bacterial biofilm.}
    \label{fig:sm}
\end{figure}

This section provides an overview of the fundamental principles underlying electrochemical communication observed in the bacterium \textit{Bacillus subtilis}.
Biofilms are complex communities of microorganisms that are surrounded by a matrix they produce themselves. As depicted in Fig. \ref{fig:sm}, a biofilm is composed of interior cells, intermediate cells, and peripheral cells. The viability and growth of bacteria within biofilms are strongly reliant upon the presence of nutrients in the surrounding fluid. As the biofilms undergo expansion, the availability of nutrients decreases for cells located in the interior regions of the biofilm. Cells located in close proximity to the peripheral of the biofilm exhibit consistent growth rates, while the interior cells experience a lack of nutrients, which can lead to their death and therefore disrupt the overall stability of the biofilm structure. In order to maintain the sustainability of the biofilm, it is crucial to establish a careful equilibrium between the rapid growth of peripheral cells and the nourishment of interior cells.

Comprehensive research has focused on resolving this complex conflict between cell growth at the biofilm periphery and the preservation of interior cells. The implications of such nutrient limitations within biofilms are discussed in \cite{liu2015}. The authors of \cite{prindle2015} proposed an intriguing mechanism involving electrochemical communication mediated by the cellular release and uptake of potassium, an essential ion for regulating cellular voltage differences.

In the experiments described by \cite{prindle2015}, glutamate, an essential nitrogen source for cell growth and maintenance, is the primary nutrient considered. Due to nutrient deficiency-induced metabolic stress, interior cells discharge potassium, resulting in hyperpolarization (more negative membrane potential). This potassium release in turn causes neighboring cells to absorb potassium and experience temporary depolarization (less negative membrane potential), disrupting their metabolic processes. As a domino effect, these neighboring cells also undergo metabolic stress and release potassium, thus further hyperpolarizing. Collectively, a potassium wave propagates from the nutrient-starved interior to the exterior of the biofilm, disrupting the pattern of nutrient consumption throughout the entire biofilm.

This potassium wave is a remarkable adaptive response that enables the biofilm to effectively redistribute nutrients. By disrupting the normal consumption pattern, the wave enables nutrient diffusion towards the starving interior cells, sustaining a moderated growth rate and preventing destabilizing levels of cell mortality in the core of the biofilm. Combining the findings of \cite{prindle2015} and \cite{liu2015} sheds light on the intricate strategies biofilms employ to surmount nutrient challenges, ensuring their continued survival and resiliency.

\section{System Model}
In this section, we focus on modeling the propagation of the potassium wave across the cross-section of a biofilm. Consequently, the analysis is done for a one-dimensional system. Given the remarkably synchronized nature of these oscillations \cite{prindle2015}, which persists even in the biofilm's remotest regions, the one-dimensional model effectively captures the biofilm's characteristics. To avoid any interference with our observations, we introduce a continuous supplement of glutamate at a rate of $\Ginput$ in the interior cells along with the input signals. This approach inhibits metabolic oscillations and, consequently, naturally occurring potassium wave propagation.

The equations representing the variation of glutamate and potassium concentrations inside and outside of cells at any location of a biofilm, as well as the equations governing membrane potential, are discussed in the following subsection.

\subsection{Extracellular Glutamate and Potassium Concentration}
The variations in the extracellular concentrations of glutamate and potassium ($\Gex$ and $\Kex$, respectively) in the biofilm are represented by 
\begin{align}
\frac{\partial \Gex}{\partial t} &=  D_G \frac{\partial^{2} \Gex}{\partial x^{2}} - \frac{\delta_{G}}{\left(1+\exp\left(V-\Vthresh\right)\right)} \Gex\left(\Gmax-\Gin\right)\nonumber\\
&\qquad+\Ginput\delta(x),\label{eq:G_ex} \\
\frac{\partial \Kex}{\partial t} &=  D_{K} \frac{\partial^{2} \Kex}{\partial x^{2}} + F g_{K} n^{4}\left(V-V_{K}\right) + F g_{L}\left(V-V_{L}\right) \nonumber\\
&\qquad- \max\left(\gamma_{K} \Kex\left(\Kmax-\Kin\right), 0\right)+\Kinput\delta(x) \label{eq:K_ex}.
\end{align}
The first terms in  \eqref{eq:G_ex} and \eqref{eq:K_ex} represent the diffusion of $\Gex$ and $\Kex$ through the exterior of the bacterial cells. The diffusion coefficients for glutamate and potassium inside the biofilm are denoted by $D_G$ and $D_K$, respectively.

The membrane potential $V$ of the cell modulates the glutamate absorption by the cell. When the membrane potential $V$ is close to the homeostatic voltage potential $\Vthresh$, this uptake is diminished. The glutamate uptake also depends on the extracellular glutamate availability and the cell's glutamate requirement, which is represented by the difference between the maximal intracellular glutamate concentration $\Gmax$ and the actual intracellular glutamate concentration $\Gin$. These variables are considered in the second element of \eqref{eq:G_ex} \cite{ford2021}. Note that, $\delta_{G}$ is the glutamate uptake rate.  The third term represents the glutamate supplement at $x=0$ (interior of biofilm).

Potassium ions are transported via potassium gates and leak gates. The potassium and leak gates with conductance $g_{K}$ and $g_L$  are controlled by a gating parameter $n$ and the potassium and leak gate reversal potentials $V_K$ and $V_L$, respectively. The uptake and release of potassium through these channels are described by the second and third term of \eqref{eq:K_ex}, which are modeled using the Hodgkin-Huxley model \cite{hodgkin1952}. The fourth term in \eqref{eq:K_ex} represents the potassium pump, which transports potassium ions from the cell's exterior to its interior \cite{ford2021}. If the internal concentration of potassium is below a certain threshold, represented by $\Kmax$, the pump will intake potassium. The extracellular potassium concentration also affects the intake. $\gamma_K$ and $F$ are potassium pump strength and voltage to potassium conversion factor, respectively. The final term with $\Kinput$ is the input signal at $x=0$.

\subsection{Boundary Conditions}

The boundary conditions for glutamate and potassium at the biofilm boundary are given by Neumann flux conditions, as follows:

\begin{align}
D_{G} \frac{\partial \Gex}{\partial x} = \frac{D_{G,\inter}}{L_b} \left(G_0-\Gintr\right) \\
D_{K} \frac{\partial \Kex}{\partial x} = \frac{D_{K,\inter}}{L_b} \left(K_0-\Kintr\right).
\end{align}

Here, the diffusion coefficients of glutamate and potassium in the fluid (in which biofilm is located), respectively, are $D_{G, \inter}$ and $D_{K, \inter}$. The concentrations of glutamate and potassium at the interface are denoted as $\Gintr$ and $\Kintr$, respectively. Meanwhile, $G_0$ and $K_0$ represent the long-range or baseline concentrations of glutamate and potassium, respectively.  The boundary layer width is represented by the parameter $L_b$.

It is important to note that at the boundary $x=0$,  $\frac{\partial \Kex}{\partial x} = 0$, means that there is no net potassium flux at the interior of the biofilm.
\subsection{Intracellular Glutamate and Potassium Concentration}

The following equations describe the dynamics of the intracellular glutamate and potassium concentrations in bacterial cells \cite{ford2021}:

\begin{align}
\frac{d \Gin}{d t} = & \frac{\delta_{G}}{\left(1+\exp\left(V-\Vthresh\right)\right)} \Gex\left(\Gmax-\Gin\right) - \gamma_{G} \Gin\left(\Mgrow + r_{b}\right), \label{eq:G_int} \\
\frac{d \Kin}{d t} = & -F g_{K} n^{4}\left(V-V_{K}\right) - F g_{L}\left(V-V_{L}\right)  \nonumber\\
&\qquad+ \max\left(\gamma_{K} \Kex\left(\Kmax-\Kin\right), 0\right), \label{eq:K_int}
\end{align}

The first term in \eqref{eq:G_int} and terms in \eqref{eq:K_int} are the same as the second term in \eqref{eq:G_ex} and second to fourth terms in \eqref{eq:K_ex}, respectively, with opposite signs. The opposite signs represent the movement across the membrane in opposite directions. The second term in \eqref{eq:G_int} represents the consumption of glutamate due to biofilm growth and metabolic activities. $\gamma_G$  denotes the glutamate consumption rate, $r_b$ denotes the glutamate used in base metabolic processes, and $\Mgrow$ denotes the propensity of the bacterial cell to grow. When the bacteria are under stress, $\Mgrow$ reduces, and the cells consume less glutamate. The value of $\Mgrow$ is given by the expression \cite{ford2021}:

\begin{align}
\Mgrow = \frac{T_G}{T_G + T_V},
\end{align}

where 
$$T_G = \frac{\Gin}{\Gin+G_u}$$
is a Hill function that becomes large when $\Gin$ is larger than the lower bound $G_u$, and $$T_V = \eta_{V} \left(\tanh\left(\gamma_{V}\left(V / V_{l} - 1\right)\right) + 1\right)$$ is a hyperbolic tangent activation function that becomes large when the membrane potential $V$ is above the bound $V_{l}$. $\eta_V$ and $\gamma_V$ are shape parameters. Biofilm grows when their $\Gin$ is high, resulting in a large $T_G$, and it is not hyperpolarized, resulting in a small $T_V$.

The  growth of the biofilm's length ($L$) is defined as follows:

\begin{align}
\frac{d L}{d t} & = \delta_{\text{g}} \int_{0}^{L} \Gin \Mgrow dx,
\end{align}

where $\delta_{\text{g}}$ is the growth rate. 

Bacteria can acclimatize to a variety of potassium concentrations over time, making potassium levels an excellent indicator of cellular stress. Let $\Kac$ denote the potassium concentration to which cells are accustomed. Then, its dynamics are governed by the equation:

\begin{align}
\frac{d \Kac}{d t} = \eta_{K}\left(\Kex-\Kac\right).
\end{align}

$\Kac$ also affects the reverse potential of the leakage gate.

\subsection{Biofilm Voltage Potential}

The set of equations representing the cell membrane voltage potential $V$ is given by:

\begin{align}
\frac{d V}{d t} = & -g_{K} n^{4}\left(V-V_{K}\right)-g_{L}\left(V-V_{L}\right)\nonumber\\
&\qquad+ \max\left(\gamma_{K} \Kex\left(\Kmax-\Kin\right), 0\right) / F ,\label{eq:Veq} \\
\frac{d n}{d t} = & \alpha \frac{\left(\Gmax-\Gin\right)^{m}}{\left(\Gmax-G_{l}\right)^{m}+\left(\Gmax-\Gin\right)^{m}}(1-n) - \beta n , \label{eq:n}\\
V_{K} = & V_{K0} + \delta_{K} \Kex, \\
V_{L} = & V_{L0} + \delta_{L}\left(\Kex-\Kac\right).
\end{align}
 The first and second terms in \eqref{eq:Veq} are from the Hodgkin-Huxley model. The third term represents the contribution of potassium dynamics to the voltage potential normalized by $F$.
The dynamics of the gating parameter $n$ are described by  \eqref{eq:n}, involving, gate opening rate $\alpha$, closing rate $\beta$, Hill coefficient $m$, glutamate concentration below which the cell hyperpolarizes $G_{l}$, $\Gmax$, and $\Gin$. When the glutamate concentration inside the cell is low ($\Gin<G_l$), the potassium gates openness increases, release potassium, and hyperpolarize. $V_K$ and $V_L$ are dependent on the concentration of potassium, the base reverse potentials $V_{K0}$ and $V_{L0}$, and the gate reversal strengths $\delta_K$ and $\delta_L$, respectively. 

When the input potassium signal is fed to the interior of the biofilm, the extracellular potassium concentration $\Kex$ increases (see equation \eqref{eq:K_ex}). The increase in $\Kex$ increases the intracellular potassium concentration $\Kin$ of the bacterium (see equation \eqref{eq:K_int}) and makes bacteria depolarized (see equation \eqref{eq:Veq}). The depolarized bacteria reduces glutamate uptake (see equation \eqref{eq:G_int}) and becomes stressed. This decrease in $\Gin$ increases $n$ and opens the potassium channel (see equation \eqref{eq:n}), releases more potassium, and hyperpolarizes. After hyperpolarization, the cells pump back the $\Kex$ inside the cell using the potassium pump after some time. The released potassium during hyperpolarization diffuses to nearby cells, and the same process continues. The intermediate cells thus amplify and relay the signal to the periphery of the biofilm. When the cell is hyperpolarized, and $\Gin$ is small, $\Mgrow$ is small and the growth of the biofilm stops. This way the input signal is propagated from the interior of the biofilm to the periphery.
\section{Numerical Analysis and Discussion}
\begin{figure}
    \centering
    \includegraphics[width=0.9\linewidth]{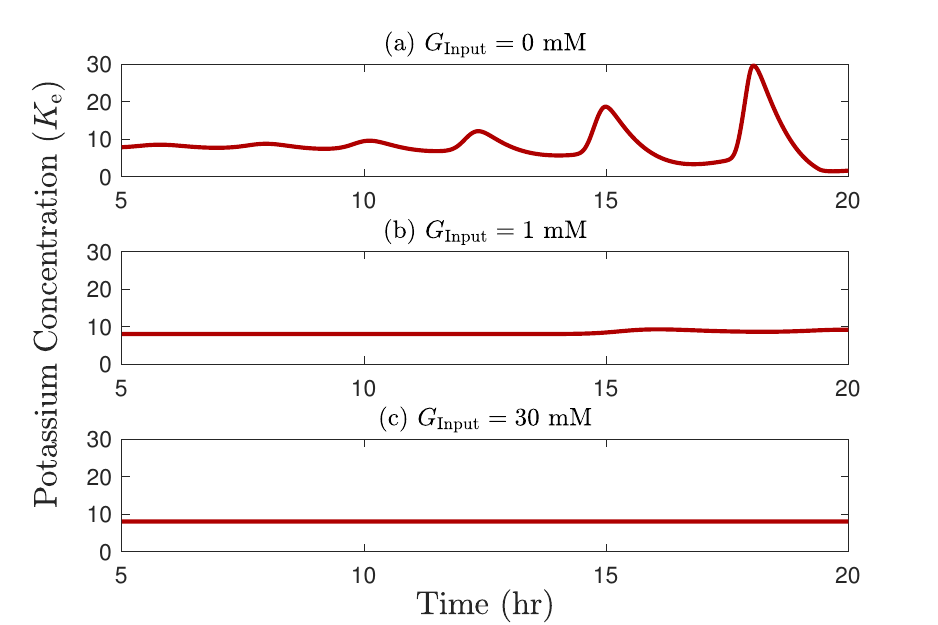}
    \caption{Potassium oscillations due to metabolic stress. Supplementing glutamate to the interior cells of biofilm quenches the oscillation.}
    \label{fig:GinvsT}
\end{figure}
We used MATLAB to evaluate the differential equations numerically with the following parameter values \cite{ford2021}: $D_{G}=0.540\ \mm^2/\hr,\ D_{K}=0.497\ \mm^2/\hr,\ \delta_G=10 \ \hr^{-1},\ F=5.6\ \mM/\mV,\ \Gmax=20 \ \mM,\ \Vthresh=-150\ \mV,\ g_K=180 \ \hr^{-1},\ g_L=1.2\ \hr^{-1},\ \gamma_K= 0.025 \left(\hr\times \mM\right)^{-1},\ \Kmax=300\ \mM,\ D_{G,\inter}=0.900\ \mm^2/\hr,\  D_{K,\inter}=4.97\ \mm^2/\hr,\ \Gintr=20\ \mM,\ \Kintr=8\ \mM,\ L_b=0.5\ \mm,\ G_0=30\ \mM,\ K_0=8\ \mM,\ \gamma_G=1.125\ \hr^{-1},\ r_b=0.1,\ G_u=18\ \mM,\ V_l=-175\ \mV,\ \eta_V=20,\ \gamma_V=20,\ \delta_g=0.0075 \mm\times \left(\mM\times \hr\right)^{-1},\ G_l=10\ \mM,\ m=2,\ V_{K0}=-380\ \mV,\ \text{and } V_{L0}=-156\ \mV$. The initial values used for the simulation are $\Gex=30\  \mM,\ \Kex=8\ \mM,\ \Gin=20\ \mM,\ \Kin=300\ \mM,\ \Kac=9\ \mM,\ V=-156\ \mV,\ L=0.12\ \mm,\ \text{and } n=0.1$. The numerical results are explained below.

The fluctuation of the extracellular potassium concentration with respect to time is depicted in Fig. \ref{fig:GinvsT}. The metabolic stress gradually builds up in the absence of glutamate at the interior of the biofilm, and the interior cells start sending potassium waves to the periphery cells to limit biofilm growth (see \ref{fig:GinvsT} (a)). We can see from Fig. \ref{fig:GinvsT} (b) and (c) that providing enough glutamate to the inner cells prevents metabolic stress and, consequently, oscillations. The observations are carried out at $x=10\ \mm$ from the interior of the biofilm.

For the remaining part of the study, glutamate is fed to the biofilm's interior cells (\ie $\ x=0$) at a rate of $30 \mM/\ms$ to avoid naturally occurring oscillations.

\begin{figure}
    \centering
    \includegraphics[width=0.85\linewidth]{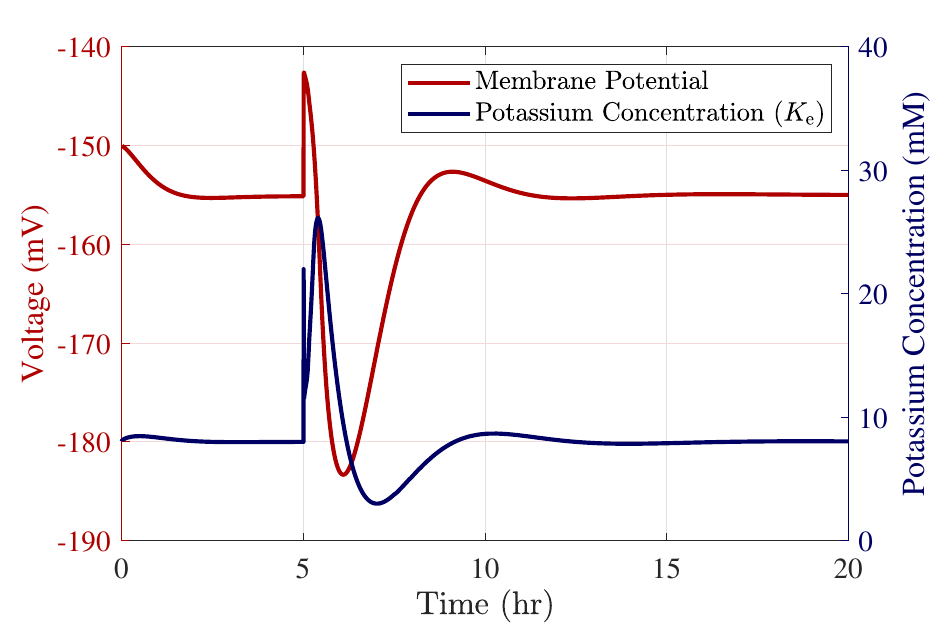}
    \caption{Channel impulse response in terms of membrane potential and extracellular potassium concentration with time when an impulse of potassium signal with magnitude $\Kinput= 100\ mM$ is given at $x=0$ at time $t=5$ hour.}
    \label{fig:KVT}
\end{figure}

\begin{figure}
    \centering
    \includegraphics[width=\linewidth]{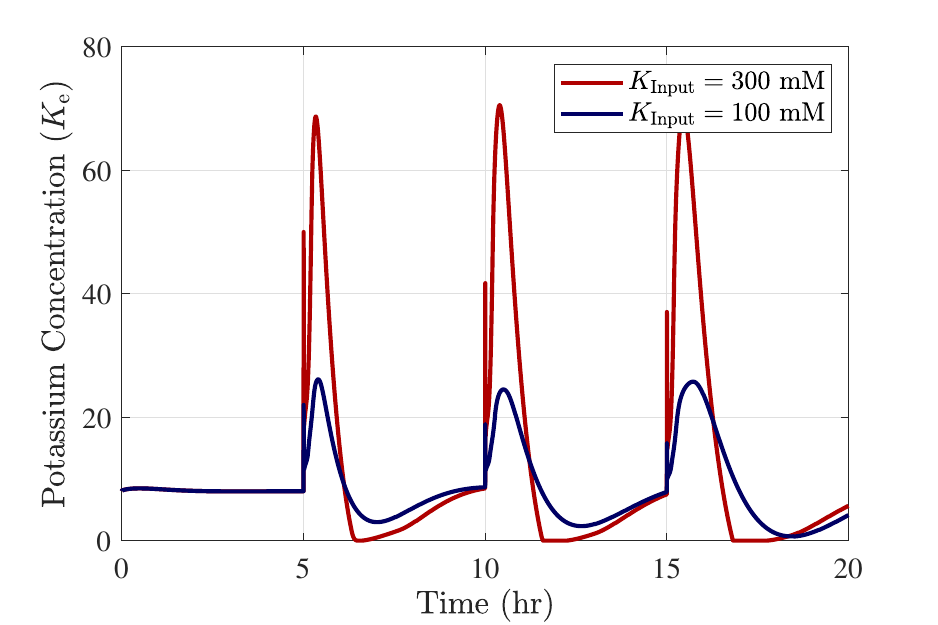}
    \caption{Extracellular potassium concentration in biofilm at $x=10 mm$ when the input to the biofilm is a series of impulses at time $t=5,\ 10$ and $15$ hour. }
    \label{fig:KinT}
\end{figure}

Fig. \ref{fig:KVT} depicts the impulse response of the channel in terms of voltage potential and extracellular potassium concentration with time when the interior cells of the biofilm are excited by an impulse emission ($\Kinput$ is an impulse signal) of potassium signal with magnitude $100\ \mM$ at time $t=5$ hours. The potassium signal is measured at distance of $10\ \mm$ from the biofilm's center.  The diffusion of potassium ions due to impulse emission 
causes the bacterial cells to absorb potassium and momentarily depolarize (less negative membrane potential). This depolarization disrupts the metabolic processes of bacterial cells, causing them to feel more stressed. These metabolically stressed bacteria release potassium (see the increase in potassium concentration in Fig. \ref{fig:KinT}) and become hyperpolarized (more negative membrane potential), which relieves metabolic stress. The potassium pump in the cell absorbs the lost potassium and restores the cell's equilibrium. Consequently, the extracellular potassium concentration and voltage potential return to a stable state. The released potassium due to hyperpolarization diffuse to nearby cells and the same process continues.
As seen in Fig. \ref{fig:KVT} and Fig. \ref{fig:KinT}, potassium's impulse emission at the biofilm's interior causes an initial spike in the extracellular potassium concentration at the observation location due to diffusion alone(not by the amplify and relay process by the intermediate cells). Our simulations, not included in the paper, confirm that the magnitude of the initial spike fades with distance. So, the short spike in concentration is due to the diffusion of the impulse signal, and the following pulse is due to hyperpolarization.

\begin{figure}
    \centering
    \includegraphics[width=0.9\linewidth]{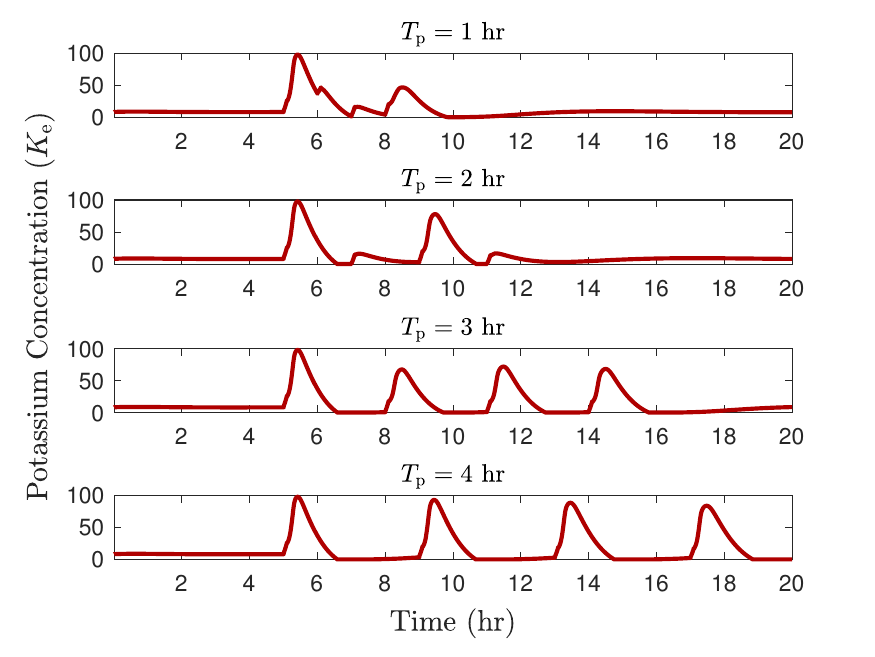}
    \caption{Variation of the extracellular potassium concentration with time with respect to the time period between input potassium pulses.}
    \label{fig:pulse}
\end{figure}

Fig. \ref{fig:KinT} illustrates the variation of extracellular potassium concentration at $x=10\ \mm$ with time due to a potassium impulse train ($\Kinput$ is an impulse train) at the interior of the biofilm at $t=5, 10$ and $15$ hour. Fig. \ref{fig:KinT} demonstrates that the impulse signals within the biofilm induce the propagation of corresponding pulse signals. Pulses are not mixed and are easily distinguishable. This is not the case when the interval between input signals is brief, as shown in Fig. \ref{fig:pulse}. Fig. \ref{fig:KinT} demonstrates that an increase in $\Kinput$ results in an increase in the number of molecules released due to hyperpolarization. 

\begin{figure}
    \centering
    \includegraphics[width=0.9\linewidth]{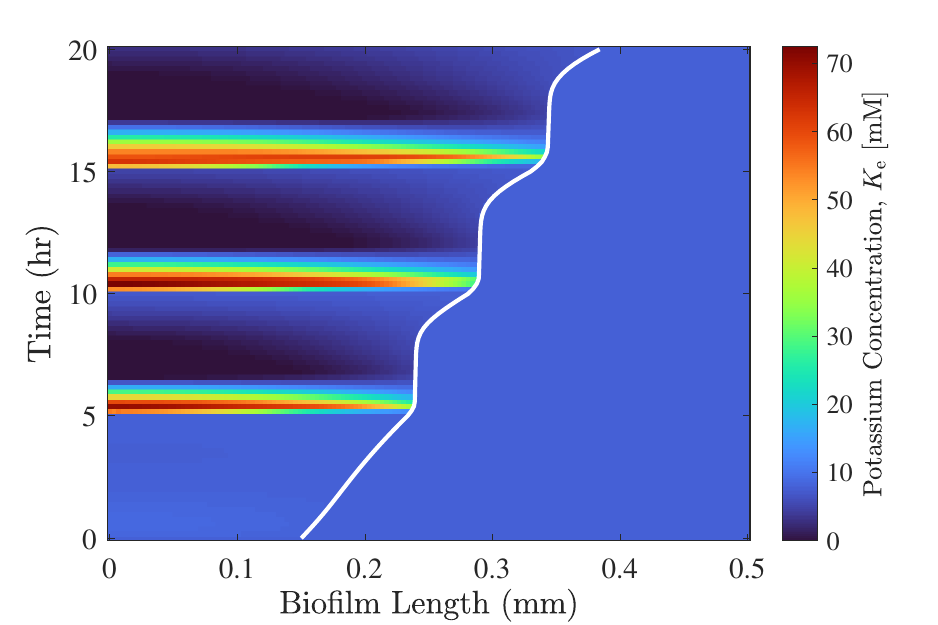}
    \caption{Variation of the extracellular potassium concentration across space and time. We can also see that potassium signals from the interior cells stop the biofilm's growth for some time. The white line represents the boundary of the biofilm.}
    \label{fig:ImpulseBio}
\end{figure}

Fig. \ref{fig:ImpulseBio} depicts the spatiotemporal variation of extracellular potassium concentration when the biofilm is stimulated with a potassium impulse train at the interior of the biofilm at times $t=5$, $10$, and $15$ hours. The white line represents the biofilm's boundary, with the left side representing the biofilm and the right side representing the fluid medium. We can verify that the potassium signal travels the length of the biofilm. Due to the release of potassium, the diffusion of the input impulse signal causes depolarization followed by hyperpolarization (red color regions). After hyperpolarization, the released potassium is pumped back into the bacterial cells, resulting in low extracellular potassium levels (dark blue regions). Also, note that this signaling has an impact on biofilm growth. When the potassium signal reaches the bacteria's peripheral cells, biofilm growth ceases.

Until now, the interior cells have been stimulated with an impulse input signal. Now, we excite the interior cells with a pulse that has a duration of $W=0.1 \ \hr$ at a rate of $5\ \mM/\ms$ potassium. Fig. \ref{fig:pulse} illustrates the variation of observed pulse shapes over time for different time intervals (denoted by $\tp$). Beginning at the fifth hour, a series of rectangular pulses of potassium signal are transmitted to interior cells. We can see that the time spread of the output signal is very large. Note that, there is no sudden initial spike of output signal  as seen in the impulse emission scenario. This is due to the low concentration and continuous emission of potassium pulse.

In contrast to diffusion-limited molecular communication, nearby pulses do not combine to produce stronger signals when the input pulses are close together. Fig. \ref{fig:pulse} demonstrates that when $\tp$ is small, the output pulses following the initial pulse diminish. When the bacteria's interior cells are stimulated by a series of potassium pulses, the bacterial cells in the biofilm undergo depolarization and then hyperpolarization during the first pulse. However, if the subsequent potassium waves are close together, the cells may not have enough potassium to release again and hyperpolarize due to a lack of time it gets to replenish the intracellular potassium by pumping the extracellular potassium that was released in the previous wave. This ultimately results in weak output signals. If $\tp$ is sufficiently high, the bacterial cells have sufficient time to replenish the potassium lost due to the previous potassium release and release enough potassium when the next signal arrives.

The above analysis confirms that stimulating the interior bacterial cells with impulse or pulse signals can produce corresponding distinguishable output signals throughout the biofilm. The input signal at the interior cells is amplified and relayed by the intermediate cells to reach the peripheral cells. In practice, the output signal can be measured by using fluorescent probes or patch clamps \cite{benarroch2020}. As emitted potassium from a biofilm diffuses towards adjacent biofilms, it can activate a similar amplify-and-relay mechanism in them. Consequently, these biofilms can function as communication nodes, paving the way for long-distance signaling in molecular communication frameworks. 

\section{Conclusions}

In this study, electrochemical communication in bacteria was investigated by introducing artificial potassium concentration signals. Our key findings include the adaptation of mathematical models to incorporate nutrient supply and various input signals, the numerical validation of continuous glutamate nutrient supply to mitigate metabolic stress and suppress natural oscillations, and the analysis of biofilm responses to impulse and pulse signals.
In addition, we have investigated the temporal aspects of input signals and demonstrated that the time interval between pulses has a substantial effect on the characteristics of output signals. This information has practical implications for the development of effective communication strategies with bacterial biofilms.
Our research highlights the potential of bacterial biofilms to serve as nodes in electrochemical communication networks. It suggests that biofilms can function as efficient information relays and synchronization points, paving the way for novel biotechnology and bioengineering applications. In our future works, we plan to validate the system through experimental investigations.

\bibliographystyle{IEEEtran}
\bibliography{Bacteria}
\end{document}